\documentclass[journal,comsoc]{IEEEtran}

\usepackage[T1]{fontenc}

\ifCLASSINFOpdf 
\usepackage[pdftex]{graphicx}
\graphicspath{{./figures/}}
\else
\usepackage[dvips]{graphicx}
\graphicspath{{./figures/}}
\fi

\ifCLASSOPTIONcompsoc
\usepackage[caption=false,font=normalsize,labelfont=sf,textfont=sf]{subfig}
\else
\usepackage[caption=false,font=footnotesize]{subfig}
\fi

\usepackage{amsmath}

\usepackage[cmintegrals]{newtxmath}

\providecommand{\expj}[1] {\ensuremath{\mathrm{e}^{\,\mathrm{j}#1}}}
\renewcommand{\expj}[1] {\ensuremath{\mathrm{e}^{\,\mathrm{j}#1}}}

\providecommand{\expmj}[1] {\ensuremath{\mathrm{e}^{-\mathrm{j}#1}}}
\renewcommand{\expmj}[1] {\ensuremath{\mathrm{e}^{-\mathrm{j}#1}}}

\providecommand*{\eu}%
{\ensuremath{\mathrm{e}}}

\providecommand*{\iu}%
{\ensuremath{\mathrm{j}}}

\makeatletter
\providecommand*{\diff}%
{\@ifnextchar^{\DIfF}{\DIfF^{}}}
\def\DIfF^#1{%
	\mathop{\mathrm{\mathstrut d}}%
	\nolimits^{#1}\gobblespace}
\def\gobblespace{%
	\futurelet\diffarg\opspace}
\def\opspace{%
	\let\DiffSpace\!%
	\ifx\diffarg(%
	\let\DiffSpace\relax
	\else
	\ifx\diffarg[%
	\let\DiffSpace\relax
	\else
	\ifx\diffarg\{%
	\let\DiffSpace\relax
	\fi\fi\fi\DiffSpace}

\begin{document}
	%
	\title{Low Peak-to-Average Power Ratio FBMC-OQAM System based on Data Mapping and DFT Precoding}
	%
	%
	%
	\author{Liming~Li, Liqin~Ding,~\IEEEmembership{Member,~IEEE,} Yang~Wang, and Jiliang~Zhang,~\IEEEmembership{Senior~Member,~IEEE}	%
		\thanks{Manuscript received XXX; revised XXX.}}

	\markboth{Journal of \LaTeX\ Class Files,~Vol.~14, No.~8, August~2015}%
	{Shell \MakeLowercase{\textit{et al.}}: Bare Demo of IEEEtran.cls for IEEE Communications Society Journals}
	%

	\maketitle
	
	\begin{abstract}
		
		Filter bank multicarrier with offset quadrature amplitude modulation (FBMC-OQAM) is an alternative to OFDM for enhanced spectrum flexible usage. To reduce the peak-to-average power ratio (PAPR), DFT spreading is usually adopted in OFDM systems. However, in FBMC-OQAM systems, because the OQAM pre-processing splits the spread data into the real and imaginary parts, the DFT spreading can result in only marginal PAPR reduction. 
		This letter proposes a novel map-DFT-spread FBMC-OQAM scheme. In this scheme, the transmitting data symbols are firstly mapped with a conjugate symmetry rule and then coded by the DFT. According to this method, the OQAM pre-processing can be avoided. Compared with the simple DFT-spread scheme, the proposed scheme achieves a better PAPR reduction. In addition, the effect of prototype filter on the PAPR is studied via numerical simulation and a trade-off exists between the PAPR and out-of-band performances. 

	\end{abstract}
	
	\begin{IEEEkeywords}
		FBMC, OQAM, DFT-Spreading, PAPR, Multipath channel.
	\end{IEEEkeywords}
	
	%
	\IEEEpeerreviewmaketitle
	
	\section{Introduction}
		
	\IEEEPARstart{F}ilter bank multicarrier (FBMC) is a candidate waveform for the mobile communications in the future. Compared with the OFDM, the FBMC employs pulse shaping by involving a prototype filter at each subcarrier to suppress the out-of-band (OOB) radiation and can provides a more flexible frequency band resourse usage owing to the low OOB radiation\cite{farhang-boroujeny_ofdm_2011}.
	To achieve the highest bandwidth efficiency while maintaining good time-frequency localization (TFL), offset quadrature amplitude modulation (OQAM) is usually adopted in FBMC systems and the orthogonality of the OQAM signal is restricted only to the real field.
	
	As a multicarrier technique, FBMC-OQAM signal has a large dynamic range in the time domain, i.e., a high peak-to average power ratio (PAPR). High PAPR will cause severe non-linear distortion if the power amplifier (PA) does not work in its linear region for power efficiency consideration. One way to cope this problem is to use a PA with digital predistortion (DPD) technology, however, this means high cost and may not be suitable for low-cost terminal devices. 
	Another way is to perform a clipping, which degrades the bit error probability (BEP)
	performance and causes spectral spreading. 

	In order to reduce the signal PAPR, a lot of baseband signal optimaizing methods have been proposed, e.g., selective mapping (SLM) and partial transmit sequence (PTS). These methods require a high computational complexity and side information (SI) and thus are difficult to be employed in practical systems. Instead, a DFT precoded OFDM system, i.e., the single carrier frequency division multiple access (SC-FDMA), is adopted in the uplink of long term evolution (LTE). 
	In SC-FDMA, a DFT spreading is performed to the data symbols prior to the OFDM modulation and can archieve the same PAPR performance as a single carrier signal while mentaining low computational complexity. In	 \cite{ihalainen_filter_2009}\cite{yuen_single_2010} \cite{bellanger_fbmc_2010}, the DFT-spreading scheme is also considered in FBMC-OQAM systems. Although simply combining the DFT-spreading and FBMC-OQAM reduces the PAPR, it does not perform as impressive as SC-FDMA. Therefore, the authors in
	\cite{ihalainen_filter_2009}\cite{bellanger_fbmc_2010} have also proposed an filter-bank-spread FBMC to provide a lower PAPR. However, the filter-bank-spread FBMC requires a higher computation and has the additional disadvantages of lower bandwidth efficiency\cite{na_low_2017}. 
	In \cite{na_low_2017}, the authors mentioned that the OQAM phase term has an influence on the PAPR performance of a simple DFT-spread FBMC system and proposed the identically-time-shifted-multicarrier (ITSM) phase term condition to reduce the PAPR.

	A key factor of that only marginal PAPR reduction can be achieved for the simple DFT-spread FBMC-OQAM is the OQAM pre-processing operation, which splits the real and imaginary parts of DFT-coded data symbols before the subcarrier modulation implemented by an inverse fast Fourier transform (IFFT) module. In order to further enhance the PAPR reduction, we propose a new DFT-spreading schme based on a symbol mapping with conjugate symmetry, which we call map-DFT-spread FBMC, to avoid the OQAM pre-processing operation. 

	%
	%
	
	\section{System Model for Simple DFT-spread FBMC-OQAM systems}
	
	\begin{figure}[!t]
		\centering
		\includegraphics[width=3.5in]{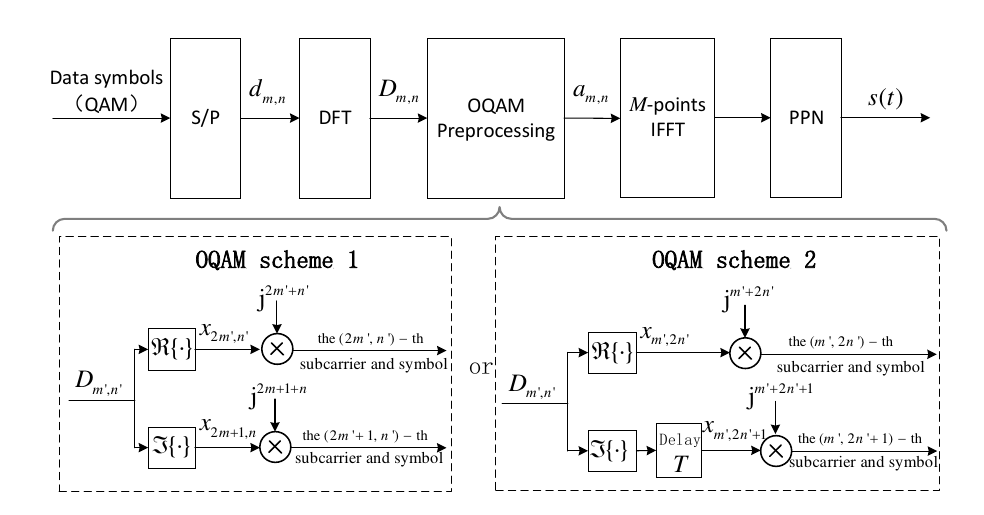}
		\caption{The simple DFT-spread FBMC-OQAM system using two kinds of OQAM pre-processing schmes}
		\label{fig_simple_dft_oqam}
	\end{figure}

\begin{figure}[!t]
	\centering
	\includegraphics[width=3.5in]{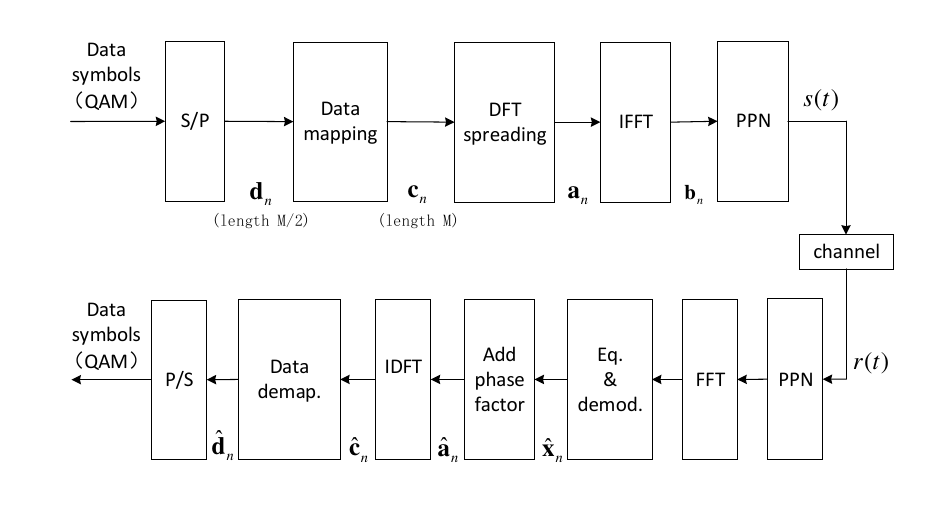}
	\caption{The baseband system architecture of the map-DFT-spread FBMC-OQAM system.}
	\label{fig_mapping_dft_fbma_oqam}
\end{figure}
	
	The	 FBMC baseband signal s(t), consisting $M$ subcarriers and $N$ symbols, can be expressed as \cite{nissel_filter_2017}:
	\begin{equation} \label{eqa_s_t}
	s(t)=\sum_{n=1}^{N}\sum_{m=1}^{M} \theta_{m,n} x_{m,n}  p(t-nT) \expj{2\pi mFt}
	\end{equation}
	
	where $(m,n)$ refers to the $m$-th subcarrier and the $n$-th symbol, $x_{m,n}$ denotes the transmitted real data symbol, $\theta_{m,n}$ is a phase factor to fulfil the real orthogonality of the OQAM signal and in most papers $\theta_{m,n} = \iu^{m+n}$, $p(t)$ denotes the prototype filter, and T and F, which satisfy $TF=1/2$, are the symbol period and subcarrier bandwidth, respectively. 
	
	In the OQAM processing, the real and imaginary parts of the original data symbol $d_{m',n'}$ are split and taken as the data value of the $x_{m,n}$ staggered in different symbol periods or different subcarriers. Condidering the different stagger patterns, there are two kinds of OQAM schemes. In the case of the simple DFT-spread FBMC-OQAM system where $d_{m',n'}$ is firstly coded by a DFT-spreading, according to the two OQAM pre-processing schemes, $x_{m,n}$ can be given by 
	\begin{flalign} 
	x_{m',2n'} = \Re \left\{ D_{m',n'} \right\} = \frac{D_{m',n'} + D_{m',n'}^{*} }{2} \label{eq_dft_type_1_x_1_D} \\ 
	x_{m',2n'+1} = \Im \left\{ D_{m',n'} \right\} = \frac{D_{m',n'} - D_{m',n'}^{*} }{2 \iu} \label{eq_dft_type_1_x_2_D}
	\end{flalign}
	for OQAM schme 1, or
	\begin{flalign} 
	x_{2m',n'} = \Re \left\{ D_{m',n'} \right\} = \frac{D_{m',n'} + D_{m',n'}^{*} }{2} \label{eq_dft_x_1_D} \\ 
	x_{2m'+1,n'} = \Im \left\{ D_{m',n'} \right\} = \frac{D_{m',n'} - D_{m',n'}^{*} }{2 \iu} \label{eq_dft_x_2_D}
	\end{flalign}
	for OQAM schme 2, where $D_{m',n'}$is the DFT of $d_{m',n'}$. Fig. \ref{fig_simple_dft_oqam} shows the system architecture of simple DFT-spread FBMC-OQAM systems.
	
	If the prototype filter is a rectangle function with effective length $T$ and only considering the signal section during $(2n+1/2)T \sim (2n+3/2)T$, $s(t)$ can be rewritten as the following \cite{na_low_2017}:
	\begin{flalign} \label{eq_dft_s_t_dft_oqam1}
	s(t) \!&=\! \frac{(-1)^{n}}{2} \! \sum_{l=0}^{M-1} \!\! \left( \! d_{\big[\!(l\!+\!\frac{M}{4}\!)\!\! \mod M \!\big] ,n} 
	\!+\!d_{\big[\!(\frac{3M}{4}\!-\!l)\!\! \mod M \!\big] ,n}^{*} \!
	\right) f_{1}\Big(\!t\!-\!\frac{l}{MF}\!\Big)
	\end{flalign}
	where $f_{1}(t)= \sum_{m=0}^{M-1} \expj{2\pi mFt}$
	can be regarded as a variant of the Dirichlet kernel function. As a comparison, in SC-FDMA systems, the output signal $s(t)$ can be written as follows:
	\begin{flalign} \label{eq_dft_s_t_sc_fdma}
	s(t) &= \sum_{l=0}^{M-1} d_{l ,n} f_{1}\Big(t-\frac{l}{MF}\Big)
	\end{flalign}
	As the Dirichlet kernel is very similar to the sinc function, the SC-FDMA output signal is similar to a single-carrier signal with sinc pulse shaping. However, in \eqref{eq_dft_s_t_dft_oqam1}, owing to the OQAM pre-processing, two data symbols are added and thus the PAPR reduction is limited\cite{na_low_2017}.	
	
	Similarly, for OQAM scheme 2, $s(t)$ can be written as
	\begin{flalign}  \label{eq_dft_s_t_dft_oqam2}
	s(t) = \frac{\iu^{n}}{2} \sum_{l=0}^{M/2-1} &  \bigg( d_{[(\frac{M}{4}+l)\!\! \mod M ] ,n} \big(1+\expj{\frac{\pi t}{T}} \big)   \nonumber \\
	& + d_{[(\frac{M}{4}-l)\!\! \mod M ] ,n}^{*} \big(1-\expj{\frac{\pi t}{T}} \big) \bigg) f_{2}\Big(t-\frac{l}{MF}\Big)  
	\end{flalign}
	where $f_{2}(t)= \sum_{m=0}^{M/2-1} \expj{4\pi mFt}$. It can also be observed that multiple data symbols are added in the output signal.
	 
\section{The map-DFT-spread FBMC-OQAM scheme}		
	As described in the previous section, the simple DFT-spread FBMC-OQAM systems cannot restore the same PAPR performance as a single-carrier signal, and this is mainly caused by the addtional OQAM pre-processing between the DFT spreading module and IFFT subcarrier modulation module. To address this problem, in this section, we introduce a new DFT-spread scheme based on a conjugate mapping opreation to avoid involving the OQAM pre-processing. 
	
	The scheme of the proposed map-DFT-spread FBMC-OQAM system is shown in Fig.\ref{fig_mapping_dft_fbma_oqam}. 
	
	In order to mentain the real orthogonality of OQAM, as described in \eqref{eqa_s_t}, a phase shift factor should be added to the transmitted real data symbols to generate a couple of staggered real and imaginary data points in the time-frequency plane. Actually, the phase factor pattern is not unique but should satisfy the following rule:
	\begin{flalign} 
	\theta_{m,n} = \pm \iu \text{ or } \pm\!1  \label{eq_dft_theta_1} \\
	\theta_{m\pm 1,n \pm\!1} = \pm \iu  \theta_{m,n} \label{eq_dft_theta_2}
	\end{flalign}
	According to the property of DFT, given a data sequence for the $n$-th FBMC-OQAM symbol, $\mathbf{a}_{n}=\mathbf{\theta}_{m,n} \circ \mathbf{x}_{m,n}$, its IDFT sequence $\mathbf{c}_{n}$ shows the following conjugate symmetry:
	\begin{equation} \label{eq_dft_c_n_csp}
	\left\{ \begin{array}{ll}
	c_{n}(0) = \lambda c_{n}^{*}(M/2)  \\
	c_{n}\left(M/2-l \right) = \lambda c_{n}^{*}(l) \\
	c_{n}\left(3M/2-l \right) = \lambda c_{n}^{*}(l)  \\
	\Im \left\{ c_{n}(M/4) \right\} = 0,\  \Im \left\{c_{n}(3M/4) \right\} = 0 \quad \text{if } \lambda = 1 \\
	\Re \left\{ c_{n}(M/4) \right\} = 0,\  \Im \left\{c_{n}(3M/4) \right\} = 0 \quad \text{if } \lambda = -1 \\
	\end{array}	\right.
	\end{equation}
	If the first element $a_{n}(0)$ is real, then $\lambda = 1$, and $c_{n}(M/4)$ and $c_{n}(3M/4)$ are real numbers. In the contrary, if the first element $a_{n}(0)$ is imaginary, then $\lambda = -1$, and $c_{n}(M/4)$ and $c_{n}(3M/4)$ are real numbers.

	If the original data sequence $\mathbf{d}_{n}$ holds the conjugate symmetry, condisering the dual operation, the DFT spreading can also directly generate the data sequence $\mathbf{a}_{n}$. For example, if the OQAM phase factor is $\theta_{m,n} = \iu^{m+n}$, we can obtain a sequence $\mathbf{c}_{n}$ with the conjugate symmetry by mapping $\mathbf{d}_{n}$ according to the following rule:
	\begin{equation} \label{eq_dft_d_c_mapping}
	c_{n}(l)=\iu^{n} \! \left\{ \!
	\begin{array}{ll}{d_{n}(0),} & \!{l=0} \\
	{d_{n}^{*}(0)} & \! {l=M / 2} \\ 
	{d_{n}(l),} & \! {l=1,2, \ldots, M / 4-1} \\ 
	{d_{n}^{*}(M / 2-l),} & \! {l=M\!/ 4+1, M\!/ 4+2, \ldots, M\!/ 2-1} \\ 
	{d_{n}(l-M\!/ 4),} & \! {l=M\!/ 2+\!1, M\!/ 2+\!2, \ldots, 3 M\!/ 4-\!1} \\ 
	{d_{n}^{*}(5 M / 4-l),} & \! {l=3 M\!/ 4+1,3 M\!/ 4+2, \ldots, M-\!1} \\ 
	{\mathfrak{R}\left\{x_{n}(M / 4)\right\},} & \! {l=M / 4} \\ 
	{\Im\left\{x_{n}(M / 4)\right\},} & \! {l=3 M / 4}
	\end{array}\right.
	\end{equation}
	
	Then, the IFFT output sequence $\mathbf{a}_{n}$ satisfies the OQAM phase requirement. Similar to \eqref{eq_dft_s_t_dft_oqam1}, considering a rectanguler filter with length $T$, $s(t)$ can be given by
	\begin{flalign} \label{eq_dft_s_t_mapping_dft_parts}
	s(t) &= \sum_{m=0}^{M-1} a_{m,n} \expj{\frac{\pi}{T}2mt} 
	= \sum_{m=0}^{M-1}\sum_{l=0}^{M-1} c_{l,n} \expmj{\frac{2\pi lm}{M}} \expj{\frac{\pi}{T}mt} \nonumber \\
	&= \sum_{l=0}^{M-1} c_{l,n} f_{1}\left(t-\frac{l}{MF} \right) F
	\end{flalign}
	
	It can be found that the proposed scheme can achieve the same PAPR performance as the single-carrier signal. However, when a filter with good TFL is adopted instead of the rectangular filter, the PAPR performance will decreased owing to the pulse shapping and symbol overlapping, and more details can be found in section \ref{sec_dft_performances}.
	
	As shown in Fig. \ref{fig_mapping_dft_fbma_oqam}, the reverse signal processing is performed at the receiver to demodulate the data symbols. 
	Firstly, the received symbol $y_{m,n}$ corresponding to $x_{m,n}(t)$ is obtained by performing analysis filtering to received signal $r(t)$:
	\begin{equation} \label{eq_dft_y_mn}	
	y_{m,n} = \int r(t) q_{m,n}(t) \, \mathrm{d}t   \\
	\end{equation}
	
	Similar to SC-FDMA, a single-tap minimum mean square error (MMSE) eqalization is adopted in the map-DFT-spread FBMC-OQAM system to conuteract the channel effect. The equalization coefficient $v_{m,n}$ at the $(m,n)$-th position is given by
	\begin{flalign} \label{eq_dft_mmse_eq_coefficients}
	v_{m,n} = \frac{(h'_{m,n})^{*}}{|h'_{m,n}|^{2}+ P_{\omega}/P_{s}} \frac{M}{\sum_{m=0}^{M-1}\frac{|h'_{m,n}|^{2}}{|h'_{m,n}|^{2}+P_{\omega}/P_{s}}}
	\end{flalign}
	where $h'_{m,n}$ is the effective channel coefficient, $P_{s}$ denotes the symbol mean power, and $P_{\omega}$ denotes the noise power. The first item on the left side of the equation \eqref{eq_dft_mmse_eq_coefficients} denotes a single-tap MMSE equalization, and the second item denotes a unbiased scale factor. 
	By taking the real part of the equalized symbol, the decision statistic ${x}'_{m,n}$ is obtained:
	\begin{flalign} \label{eq_dft_x_mn_receiving}
	x'_{m,n} = \Re \{y_{m,n} v_{m,n}\}
	\end{flalign}
	After the symbol decision operation, the estimation $\hat{x}_{m,n}$ is obtained .
	
	To obtain the original transmitted data, the OQAM phase factor need to be added to $\hat{x}_{m,n}$ to formulate $\hat{a}_{m,n}$:
	\begin{flalign} \label{eq_dft_a_mn_receiving}
	\hat{a}_{m,n} = \iu^{m+n} \hat{x}_{m,n}
	\end{flalign}
	Then, the received data $\hat{d}_{n}(l)$ can be obtained by demapping the $\hat{c}_{m,n}$, which is the IDFT of $\hat{a}_{m,n}$. According to \eqref{eq_dft_d_c_mapping}, the demapping can be performed as follows:
	\begin{flalign} \label{eq_dft_c_d_demapping}
	\hat{d}_{n}(l) \!=\! \left\{\begin{array}{l}{\! 
		\left(\hat{c}_{n}(0)+\hat{c}_{n}^{*}\big(\frac{M}{2}\big)\right) / 2, \quad \text{for }l=0} \\ {\!\left(\hat{c}_{n}(l)+\hat{c}_{n}\big(\frac{M}{2}-l\big)\right) / 2, \quad \text{for }l=1,2, \ldots, \frac{M}{4}-1} \\
	{\!\Re\left\{\hat{c}_{n}\big(\frac{M}{4}\big)\right\}+j \Im\left\{\hat{c}_{n}\big(\frac{3M}{4}\big)\right\} \quad \text{for }l=\frac{M}{4}} \\ 
	{\! \left(\hat{c}_{n}\big(l \!\!+\!\! \frac{M}{4}\big) \!\!+\!\! \hat{c}_{n}\big(\frac{5M}{4} \!\!-\!\! l\big)\right) / 2, \quad \text{for }l \!=\! \frac{M}{4} \!\!+\!\! 1, \frac{M}{4} \!\!+\!\! 2, \ldots \frac{M}{2} \!\!-\!\! 1}\end{array}\right.
	\end{flalign}
	
	\begin{figure*}[!t]
	\centering
	\subfloat[4-QAM]{\includegraphics[width=2.4in]{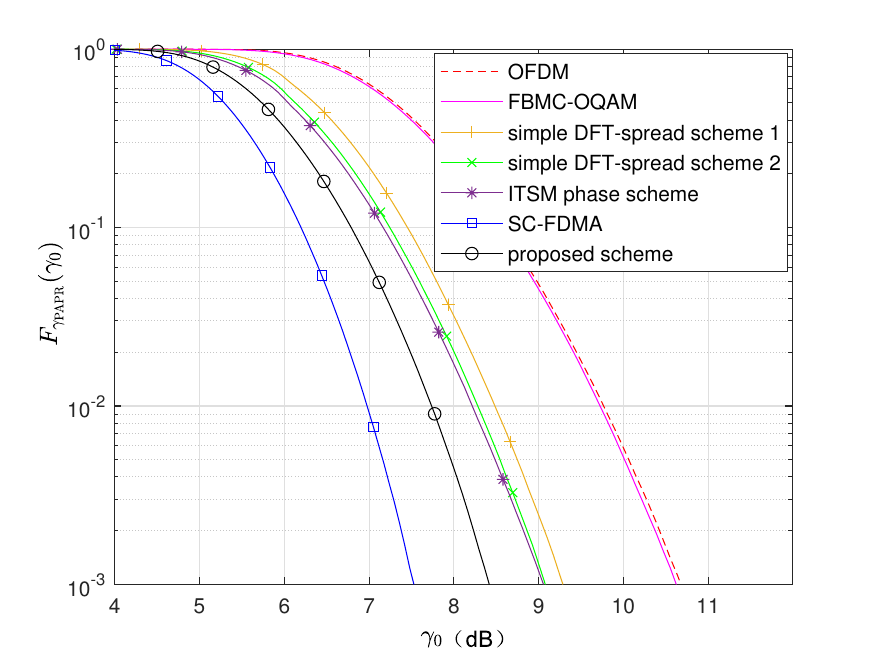}
		\label{fig_papr_4pam}} 
	\subfloat[4-QAM with different filters]{\includegraphics[width=2.4in]{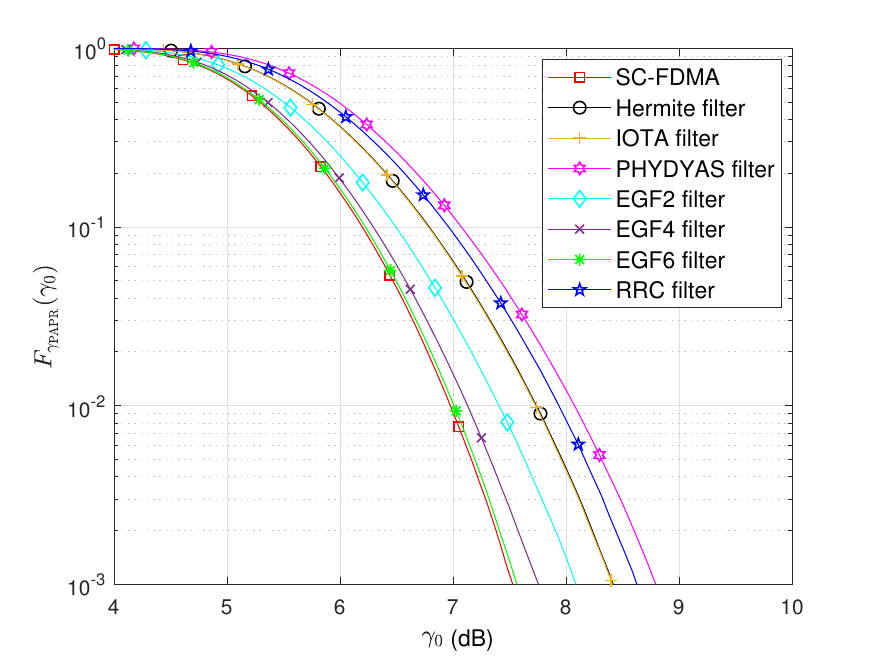}
		\label{fig_papr_4pam_filters}} 
	\subfloat[The power spectrum density curves.]{\includegraphics[width=2.4in]{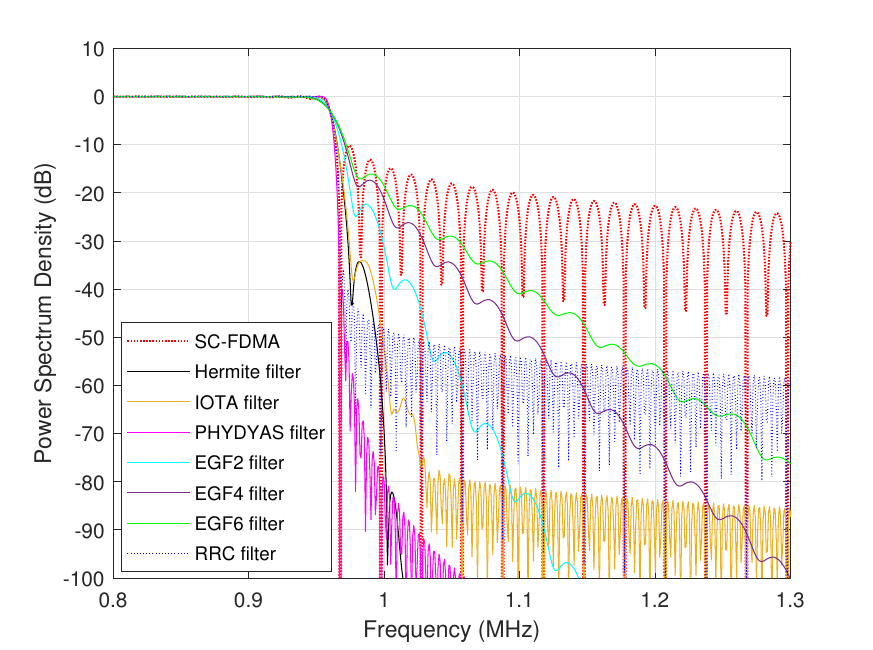}
		\label{fig_dft_papr_psd_filters}} 
	\caption{The PAPR and OOB performances of different DFT spreading schemes.}
	\label{fig_papr_performances}
	\end{figure*}
	
	\begin{figure*}[!t]
		\centering
		\subfloat[AWGN]{\includegraphics[width=2.4in]{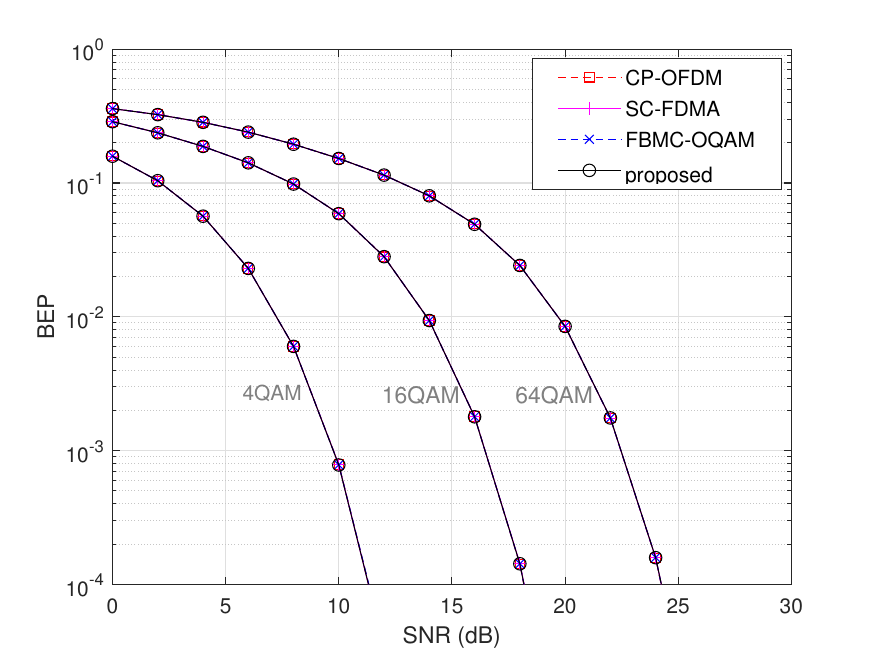}
			\label{fig_bep_awgn}} 
		\subfloat[Pedestrian A]{\includegraphics[width=2.4in]{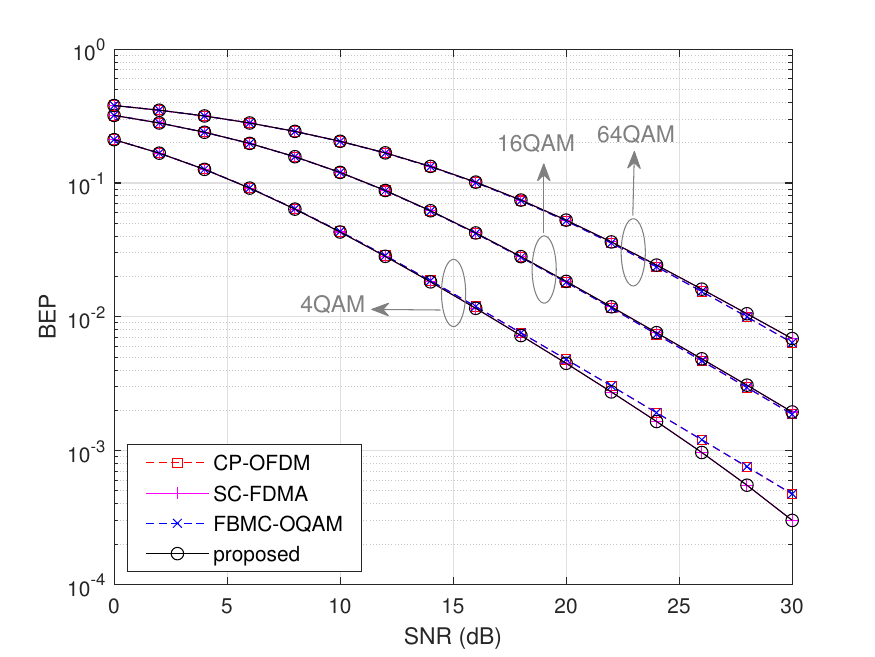}
			\label{fig_bep_peda}} 
		\subfloat[Vehicular A]{\includegraphics[width=2.4in]{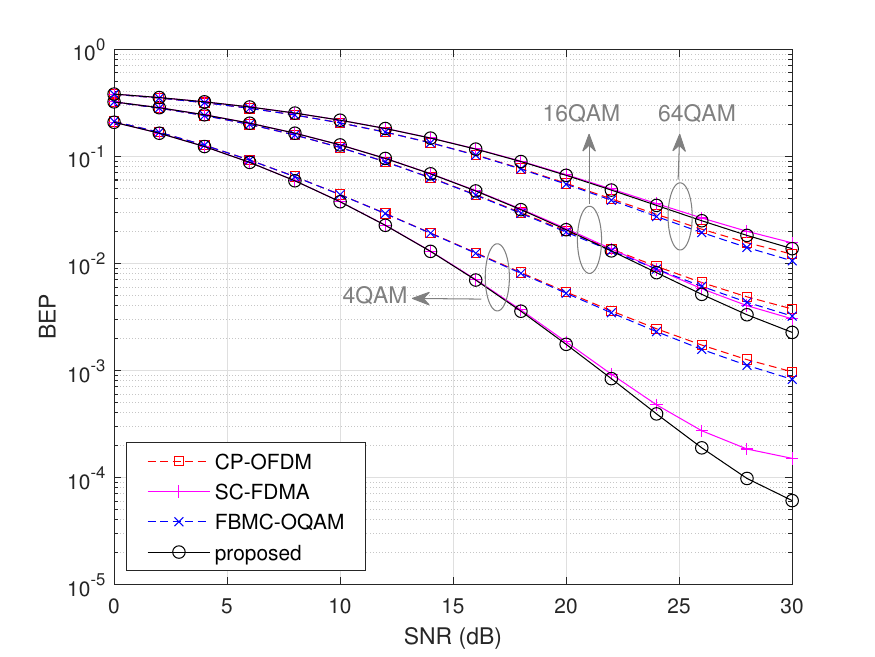}
			\label{fig_bep_veha}} 
		\caption{The BEP performances over AWGN and multipath channels.}
		\label{fig_bep_performances}
	\end{figure*}

	\section{Simulation} \label{sec_dft_performances}
	This section evaluates the PAPR and BEP performances of the map-DFT-spread FBMC-OQAM system. The subcarrier number $M$ is set as 64, and the subcarrier bandwidth is 15KHz. The PAPR denoted by $\gamma_{\text{PAPR}}$ is calculated as follows:
	\begin{flalign} \label{eq_dft_papr}
	\gamma_{\text{PAPR}} = \frac{\max |s(t)|^{2}}{\mathbb{E} \left[ |s(t)|^{2} \right] }
	\end{flalign}
	In this paper, we use the following complementary cumulative distribution function (CCDF), $F_{\gamma_{\text{PAPR}}}(\gamma_{0})$, to evaluate the PAPR performance: 
	\begin{flalign} \label{eq_dft_ccdf}
	F_{\gamma_{\text{PAPR}}}(\gamma_{0}) = \Pr \left( \gamma_{\text{PAPR}} \geq \gamma_{0} \right)
	\end{flalign}
	where $\gamma_{0}$ denotes a threshold value. 
	
	In practical FBMC-OQAM systems, the pulse shaping is performed using a filter with good TFL instead of the rectangular filter which is discussed in the previous section. The pulse shaping and symbol overlapping will affect the PAPR performance. Fig. \ref{fig_papr_4pam} shows the CCDF curves of different systems, where FBMC systems use the Hermite filter with an overlapping factor 4. For instance, at CCDF of $10^{-3}$ The simple DFT-spread scheme 1 and 2 can reduce the PAPR about 1.5dB and 1.3dB, repectively, compared with the original FBMC-OQAM. The proposed map-DFT-spread scheme archieves a PAPR reduction around 2.2dB which is better than the simple DFT-spread and ITSM phase schemes. 
	
	However, the PAPR of map-DFT-spread FBMC-OQAM is still not as good as SC-FDMA, and this is mainly due to the pulse shaping and symbol overlapping caused by filtering, which means that different filters will cause different PAPR performances. We compare several filters in Fig. \ref{fig_papr_4pam_filters}, the PHYDYAS, IOTA, RRC and EGF filters\cite{sahin_survey_2014}. As shown in Fig. \ref{fig_papr_4pam_filters}, the EGF filter with a spreading parameter $\alpha = 6$, we call it the EGF6 filter in this paper, can archieve similar PAPR performance as SC-FDMA, however, the OOB is increased, which can be observed in Fig. \ref{fig_dft_papr_psd_filters}. A trade-off should be considered between the PAPR and OOB performances. 
	
	The BEP performance of the map-DFT-spread FBMC-OQAM is also evaluated. As shown in Fig. \ref{fig_bep_awgn}, the map-DFT-spread scheme performs the same BEP as original FBMC in the AWGN channel, which means the proposed scheme holds good orthogonality. Whereas in the case of multipath channels, the two DFT-spread schemes, the proposed map-DFT-spread scheme and SC-FDMA, show better BEP performances than the original FBMC and OFDM systems, which means a frequency diversity gain is achieved. In the case of 4QAM data symbols, the BEP performance gap is more obvious in the Vechicular A channel which has a more harsh time-frequency selectivity, and the map-DFT-spread scheme shows a better BEP than the SC-FDMA owing to that FBMC performs better than OFDM in highly time-varying channels\cite{nissel_filter_2017}. However, for 64QAM symbols, when the ISI and ICI is large, the interference among data symbols is increased owing to the spreading, and thus a more complicated equalization scheme should be considered to improve the BEP performance. 

	\section{Conclusion}
	In this paper, the map-DFT-spread FBMC-OQAM is proposed to improve the PAPR performance. Instead of the OQAM pre-processing, the data symbols with OQAM phase factors are generated directly by the combination of the conjugate symmetric mapping and the DFT spreading, and the map-DFT-spread scheme can achieve a similar PAPR as the single-carrier signal when a rectangular filter is used. However, the pulse shaping in FBMC systems limits the PAPR reduction, and a trade-off should be considered for the PAPR and OOB performances. Compared with the simple DFT-spread scheme and ITSM conditioned phase scheme, the proposed scheme can achieve a better PAPR performance while maintaining low computation complexity. The proposed scheme also benefits from the frequency diversity provided by the DFT spreading and thus can achieve good BEP performance for low-order QAM symbols over multipath fading channels. 
	
	
	
	%

	%
	%
	%
	%
	%

	\ifCLASSOPTIONcaptionsoff
	\newpage
	\fi

	
	
	\bibliographystyle{IEEEtran}
	\bibliography{IEEEabrv,F:/project/zotero/lib/ee_communication}

\begin{thebibliography}{1}
\providecommand{\url}[1]{#1}
\csname url@samestyle\endcsname
\providecommand{\newblock}{\relax}
\providecommand{\bibinfo}[2]{#2}
\providecommand{\BIBentrySTDinterwordspacing}{\spaceskip=0pt\relax}
\providecommand{\BIBentryALTinterwordstretchfactor}{4}
\providecommand{\BIBentryALTinterwordspacing}{\spaceskip=\fontdimen2\font plus
\BIBentryALTinterwordstretchfactor\fontdimen3\font minus
  \fontdimen4\font\relax}
\providecommand{\BIBforeignlanguage}[2]{{%
\expandafter\ifx\csname l@#1\endcsname\relax
\typeout{** WARNING: IEEEtran.bst: No hyphenation pattern has been}%
\typeout{** loaded for the language `#1'. Using the pattern for}%
\typeout{** the default language instead.}%
\else
\language=\csname l@#1\endcsname
\fi
#2}}
\providecommand{\BIBdecl}{\relax}
\BIBdecl

\bibitem{farhang-boroujeny_ofdm_2011}
B.~Farhang-Boroujeny, ``{OFDM} versus filter bank multicarrier,'' \emph{IEEE
  Signal Processing Magazine}, vol.~28, no.~3, pp. 92--112, 2011.

\bibitem{ihalainen_filter_2009}
T.~Ihalainen, A.~Viholainen, T.~H. Stitz, M.~Renfors, and M.~Bellanger,
  ``Filter bank based multi-mode multiple access scheme for wireless uplink,''
  in \emph{2009 17th {European} {Signal} {Processing} {Conference}}, Aug. 2009,
  pp. 1354--1358.

\bibitem{yuen_single_2010}
C.~H. Yuen, P.~Amini, and B.~Farhang-Boroujeny, ``Single carrier frequency
  division multiple access ({SC}-{FDMA}) for filter bank multicarrier
  communication systems,'' in \emph{2010 {Proceedings} of the {Fifth}
  {International} {Conference} on {Cognitive} {Radio} {Oriented} {Wireless}
  {Networks} and {Communications}}, Jun. 2010, pp. 1--5.

\bibitem{bellanger_fbmc_2010}
M.~Bellanger, D.~Le~Ruyet, D.~Roviras, M.~Terré, J.~Nossek, L.~Baltar, Q.~Bai,
  D.~Waldhauser, M.~Renfors, and T.~Ihalainen, ``{FBMC} physical layer: a
  primer,'' \emph{PHYDYAS, January}, vol.~25, no.~4, pp. 7--10, 2010.

\bibitem{na_low_2017}
D.~Na and K.~Choi, ``Low {PAPR} {FBMC},'' \emph{IEEE Transactions on Wireless
  Communications}, vol.~PP, no.~99, pp. 1--1, 2017.

\bibitem{nissel_filter_2017}
R.~Nissel, S.~Schwarz, and M.~Rupp, ``Filter bank multicarrier modulation
  schemes for future mobile communications,'' \emph{IEEE Journal on Selected
  Areas in Communications}, vol.~35, no.~8, pp. 1768--1782, Aug. 2017.

\bibitem{sahin_survey_2014}
A.~Sahin, I.~Guvenc, and H.~Arslan, ``A survey on multicarrier communications:
  prototype filters, lattice structures, and implementation aspects,''
  \emph{IEEE Communications Surveys \& Tutorials}, vol.~16, no.~3, pp.
  1312--1338, 2014.

\end{thebibliography}
\end{document}